# The AliEn system, status and perspectives

P. Buncic
*Institut für Kernphysik, August-Euler-Str. 6, D-60486 Frankfurt, Germany and*
*CERN, 1211, Geneva 23, Switzerland*
A. J. Peters, P.Saiz
CERN, 1211, Geneva 23, Switzerland

In preparation of the experiment at CERN's Large Hadron Collider (LHC), the ALICE collaboration has developed AliEn, a production environment that implements several components of the Grid paradigm needed to simulate, reconstruct and analyse data in a distributed way. Thanks to AliEn, the computing resources of a Virtual Organization can be seen and used as a single entity – any available node can execute jobs and access distributed datasets in a fully transparent way, wherever in the world a file or node might be. The system is built around Open Source components, uses the Web Services model and standard network protocols to implement the computing platform that is currently being used to produce and analyse Monte Carlo data at over 30 sites on four continents. Several other HEP experiments as well as medical projects (EU MammoGRID, INFN GP-CALMA) have expressed their interest in AliEn or some components of it. As progress is made in the definition of Grid standards and interoperability, our aim is to interface AliEn to emerging products from both Europe and the US. In particular, it is our intention to make AliEn services compatible with the Open Grid Services Architecture (OGSA). The aim of this paper is to present the current AliEn architecture and outline its future developments in the light of emerging standards.

## 1. INTRODUCTION

### 1.1. ALICE Experiment at CERN LHC

ALICE [1] is one of the four LHC (Large Hadron Collider) experiments, currently being built at CERN, Geneva. When the experiment starts running, it will collect data at a rate of up to 2PB per year and probably run for 20 years while generating more than $10^9$ data files per year in more than 50 locations worldwide.

AliEn [2] (ALIce ENvironment) is a distributed computing environment developed by the ALICE Offline Project with the aim to offer the ALICE user community a transparent access to worldwide distributed computing and storage resources. The intention is to provide a functional computing environment that fulfils the needs of the experiment in the preparation phase and, at the same time, defines a stable interface to the end users that will remain in place for a long time, shielding the ALICE core software from inevitable changes in the technologies that make distributed computing possible. Nowadays, such technologies are commonly associated with the Grid computing paradigm.

Interfacing to common Grid solutions [3] has always been one of the primary design goals of AliEn and remains the top priority in continuing development efforts. In addition, AliEn provides its native, fully functional Grid environment based on a Web Services model. It has been put into production for ALICE users at the end of 2001. AliEn is meant to be non-intrusive, adaptive and easy to deploy in the standard environment of large computing centres typically found in HEP environment.

Currently, ALICE is using AliEn for the distributed production of Monte-Carlo data, detector simulation and reconstruction at more than 30 sites located on four continents.

### 1.2. ALICE Computing Model

The core of the ALICE Computing Model is AliRoot [4], an Object Oriented framework written in C++. It uses directly the ROOT [5] framework for performance and simplicity reasons. ROOT provides data persistency on the file level, an interface to various utility libraries, visualization, a graphical user interface, virtual Monte Carlo and a geometrical modeller. This approach allows for fast prototyping and development by using local files containing one or more events, which is very much appreciated by ALICE users and developers. The user code (simulation of detector components and reconstruction of the events) as well as the analysis code, which operates on the output of the reconstruction step, has to be linked only with ROOT libraries. In that spirit, using the AliEn C/C++ API, we have extended the capabilities of ROOT by providing an AliEn specific implementation for an abstract TGrid class to allow ALICE users a transparent access to datasets on the Grid as if they were local files. From the ROOT prompt, users can authenticate themselves, access distributed datasets or request the execution of their algorithms across distributed datasets and retrieve the resulting objects in an interactive or batch session. The system takes care of job splitting and execution and tries to optimize network traffic. When required, the datasets can be replicated and cached. The result of each job is optionally validated and the final output from many concurrent jobs is merged together and presented to the user as a dataset in his portion (directory) of the global logical file namespace (AliEn File Catalogue). The system keeps track of the basic provenance of each executed job or file transfer.

### 1.3. Use Cases

In a typical HEP experiment, during preparation and running phase, large-scale simulation must be carried out involving a large portion of the available computing resources. The simulation of ALICE events has a few specific requirements (it takes up to 24h on a 1GHz CPU to perform a detailed simulation of the detector response and the resulting output file is up to 2 GB) but the overall requirements are compatible with the use cases of other LHC experiments as described in the HEPCAL [6] document. In broad terms, these use cases deal with simulation and reconstruction, event mixing and analysis and all of them are addressed in the AliEn Grid implementation. With the exception of the analysis use case, which for ALICE is obviously bound to the ROOT framework, a large-scale simulation and reconstruction can be setup by using AliEn without dependency on ROOT.

## 2. ALIEN ARCHITECTURE

### 2.1. Building AliEn

AliEn has been built from a large number of Open Source components, re-using their functionality without modifying them. All required third party packages can be installed by using the Pacman [7] tool from the cache which can be found at http://alien.cern.ch/cache. The installation does not require root privileges and is entirely self-contained. If not already installed, the AliEn installation procedure will automatically install required third party components. At present, Intel i386 and ia64 platforms are supported under RedHat Linux 6.1 and higher.

The binary distributions for supported platforms can be found at http://alien.cern.ch/dist. They are split into packages (RPMs and tar files) according to the functionality they provide. In the simplest case (client access to an AliEn service) the end user has to install only the Base and Client packages. Other available packages are: Server, Portal, Monitor, CE (Computing Element), SE (Storage Element) and the packages specific for a Virtual Organization.

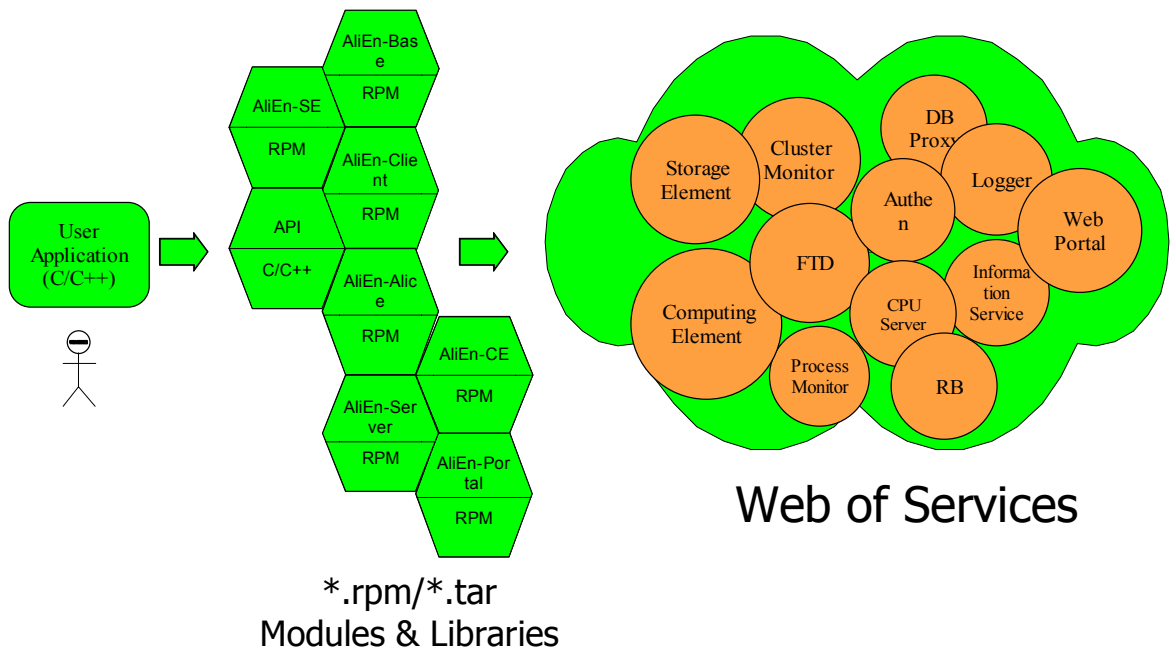

Figure 1: AliEn Components and Services

Once the components are installed and a Virtual Organization is configured, a number of AliEn Web Services at central and remote sites must be started in order to create a Web of collaborating services that together constitute the AliEn Grid.

Out of a total of 3M lines of code, only 1% corresponds to native AliEn code while 99% of the code has been imported in form of Open Source components. This made the development fast and allowed the preliminary version of the system to be in production only six months after the start of its developments.

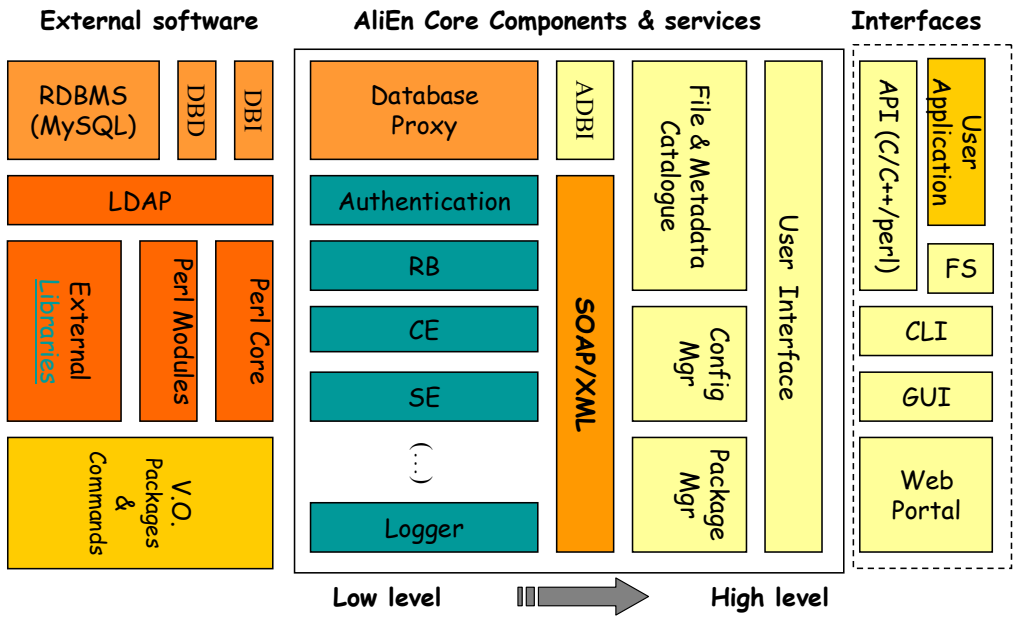

Figure 2: The architecture of AliEn - building blocks

The building blocks of AliEn can be grouped into three major categories: low-level external software components, AliEn core components and services and high-level user interfaces. In the following section the details of the components in each of these categories will be presented.

### 2.2. External Software

#### 2.2.1. Perl Core and Modules

The principal reason for using the perl [8] scripting language has been the availability of a large number of re-usable Open Source modules [9] which provide very complete cryptography support, implement a full featured SOAP [10] client and server platform and offer an easy integration with the Web for control and reporting. At present, AliEn is using around 170 external components, most of them perl modules but also third party Open Source packages and libraries on which these modules depend.

In particular, perl has a flexible database access module allowing applications to use an abstract Database Interface (DBI) and to specify a Database Driver (DBD) at run time.

#### 2.2.2. LDAP

AliEn uses a hierarchical database (LDAP – Lightweight Directory Access Protocol) to describe the static configuration for each Virtual Organization (VO). This includes *People, Roles, Packages, Sites* and *Grid Partitions* as well as the description and configuration of all services on remote sites. The code that is deployed on remote sites or user workstations does not require any specific VO configuration files, everything is retrieved from the LDAP configuration server at run time thus allowing user to select VO dynamically. At present, there are 10 AliEn VOs configured, each one effectively implementing a standalone Grid and running its own set of supporting services.

### 2.2.3. VO Packages & Commands

Each VO has to provide the description of its specific software in order to make it available to the AliEn Package Manager (See 2.3.5). This can be done by extending the default *Command* and *Package* classes in AliEn and requires a description of all commands that users can run across the Grid for this VO. The *Commands* can depend on *Packages* and *Packages* can depend on other *Packages*. Each *Package* knows how to prepare its run time environment and can carry out special steps before job start as well as upon job completion (to register job output, for example). Each *Command* may have an associated validation procedure which will verify that the command actually completed without error.

## 2.3. AliEn Core Components

### 2.3.1. Database Interface

The backend of AliEn is a relational database or, in a more general case, a federation of relational databases. While this, in principle, can also be a federation of heterogeneous databases in practice we use the MySQL databases. To avoid linking with any specific database driver on the client side, all connections to the databases are channelled via a Proxy Service. The application connects to the Proxy Service by means of a special AliEnProxy driver so that the real database driver and libraries need to be installed only in the place where an instance of the database Proxy Service is running. The Proxy Service also acts as an Authentication Service for a certificate-based authentication (see 2.4.1).

### 2.3.2. File Catalogue

The File Catalogue was the first component which has been developed and was the initial goal of the project. The complete distributed computing environment came later as a natural extension of the "file system" paradigm once we realized how easily this can be done once the basic building blocks are under control. Unlike *real* file systems, the File Catalogue does not own the files; it only keeps an association between the Logical File Name (LFN) and (possibly more than one) Physical File Names (PFN) on a real file or mass storage system. PFNs describe the physical location of the files and include the access protocol (rfio, rootd), the name of the AliEn Storage Element and the path to the local file. The system supports file replication and caching and will use this information when it comes to scheduling jobs for execution. The directories and files in the File Catalogue have privileges for owner, group and the rest of the world. This means that every user can have exclusive read and write privileges for his portion of the logical file namespace (home directory).

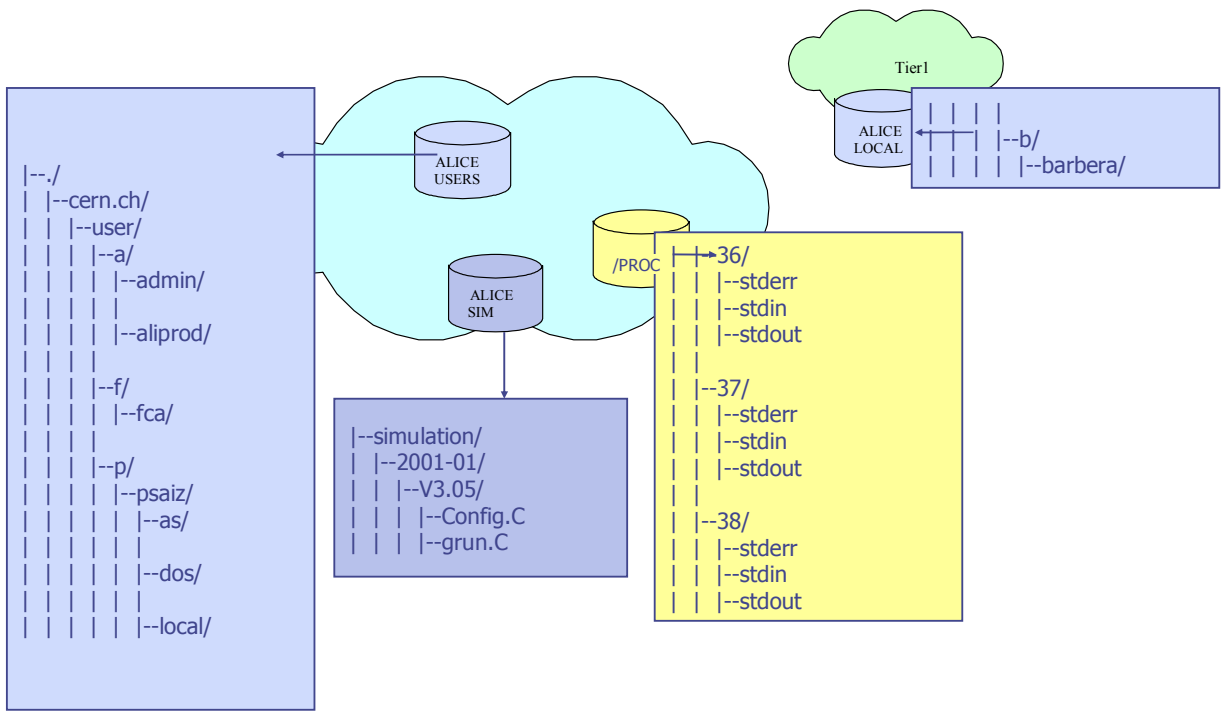

Figure 3: The hierarchy of the AliEn File Catalogue

In order to address the problem of scalability, the AliEn File Catalogue is designed to allow each directory node in the hierarchy to be supported by different database engines, possibly running on different host and, in future version, even having different internal table structures, optimized for a particular branch.

The File Catalogue is not meant to support only regular files – we have extended the file system paradigm and included information about running processes in the system (in analogy with the /proc directory on Linux systems). Each job sent to AliEn for execution gets an unique *id* and a corresponding */proc/id* directory where it can register temporary files, standard input and output as well as all job products. In a typical production scenario, only after a separate process has verified the output, the job products will be renamed and registered in their final destination in the File Catalogue.

The entries (LFNs) in the AliEn File Catalogue have an immutable unique file id attribute that is required to support long references (for instance in ROOT) and symbolic links.

### 2.3.3. Metadata Catalogue

The hierarchy of files and directories in the AliEn File Catalogue reflects the structure of the underlying database tables. In the simplest and default case, a new table is associated with each directory. In analogy to a file system, the directory table can contain entries that represent the files or again subdirectories. Due to this internal structure, it is possible to attach to a given directory table an arbitrary number of additional tables, each one having a different structure and possibly different access rights while containing metadata information that further describes the content of files in a given directory. This scheme is highly granular and allows fine access control. Moreover, if similar files are always catalogued together in the same directory this can substantially reduce the amount of metadata that needs to be stored in the database. In the example below, the search will first select all tables on the basis of the file name selection and then locate all tables that correspond to a tag definition, apply the selection and finally return only the list of LFNs for which the attribute search has been successful:

lfn:///alice/sim/2001*/V3.05/%/*.root?MonteCarlo:npart>100

While having to search over a potentially large number of tables may seem ineffective, the overall search scope has been greatly reduced using the file system hierarchy paradigm and, if data are sensibly clustered and directories are spread over multiple database servers, we could even execute searches in parallel and effectively gain performance while assuring scalability.

#### 2.3.4. Configuration Manager

The Configuration Manager is responsible for discovery and read-only interactions with the LDAP server (Section 2.2.2). It extracts all relevant configuration parameters that apply to a particular VO in the context of a site and a specific host. The information is kept in a cache to avoid frequent LDAP lookups.

#### 2.3.5. Package Manager

As mentioned in Section 2.2.3, each VO can provide the *Packages* and *Commands* that can be subsequently executed on AliEn Grid. Once the corresponding tar files with bundled executables and libraries are published in the File Catalogue and registered in a LDAP directory, the AliEn Package Manager will install them automatically as soon as a job becomes eligible to run on a site whose policy accepts these jobs. While installing the package in a shared package repository, the Package Manager will resolve the dependencies on other packages and, taking into account package versions, install them as well. This means that old versions of packages can be safely removed from the shared repository and, if these are needed again at some point later, they will be re-installed automatically by the system. This provides a convenient and automated way to distribute the experiment specific software across the Grid and assures accountability in the long term.

### 2.4. AliEn Services

AliEn Services play the central role in enabling AliEn as a distributed computing environment. The user interacts with them by exchanging SOAP messages and they constantly exchange messages between themselves behaving like a true Web of collaborating services. In the following section, the most important AliEn services will be described.

#### 2.4.1. Authentication

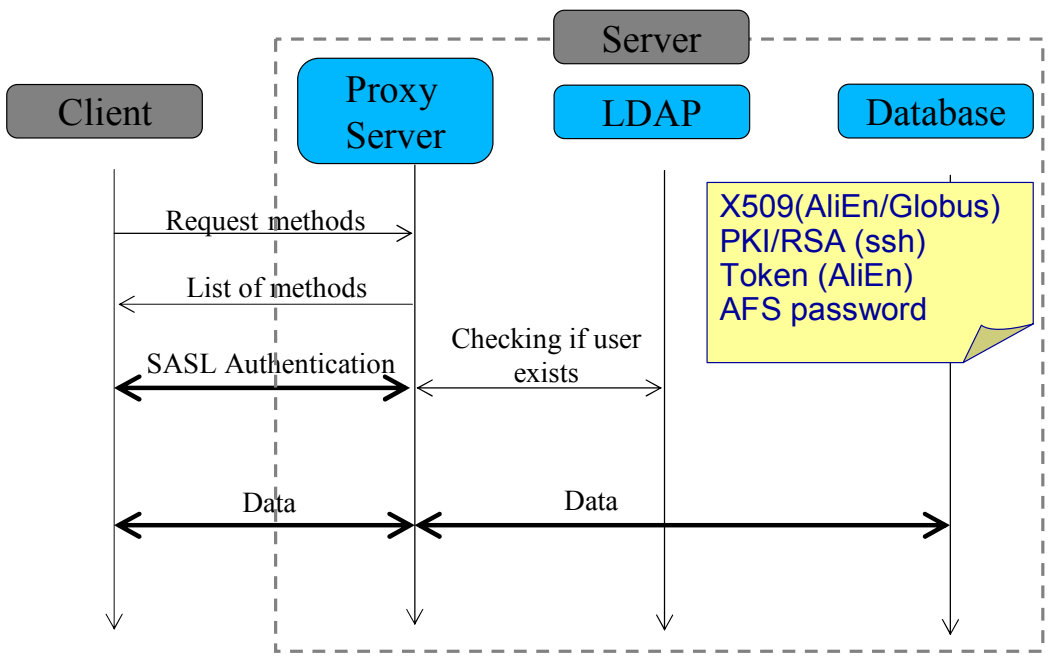

Figure 4: The sequence diagram of AliEn authentication

The Authentication Service is responsible for checking user's credentials. AliEn uses the SASL [11] protocol for authentication and implements several SASL mechanisms (GSSAPI using Globus/GSI, AFS password, SSH key, X509 certificates and AliEn tokens)

Upon successful authentication a Proxy Service acquires and holds the real database handle on behalf of a user and returns a temporary access token which the user has to present in order to re-connect to the database. The token remains in user possession and is valid for a limited period of time.

#### 2.4.2. Cluster Monitor

The Cluster Monitor service runs on a remote site and acts as a gatekeeper. It can interact with other services and provides proxy functionality for services that are behind the firewall or on a private network.

#### 2.4.3. Resource Brokers

AliEn jobs use the Condor ClassAds [12] as a Job Description Language (JDL). When users submit a job to the system it enters into the task queue and the Resource Broker [13] becomes responsible for it. The Broker analyses job requirements in terms of requested input files, requirements on execution nodes, and job output. The JDL defines the executable, its arguments and the software packages or data that are required by the job. The Broker can modify the job's JDL entry by adding or elaborating requirements based on the detailed information it can get from the system like the exact location of the dataset and replicas, client and service capabilities.

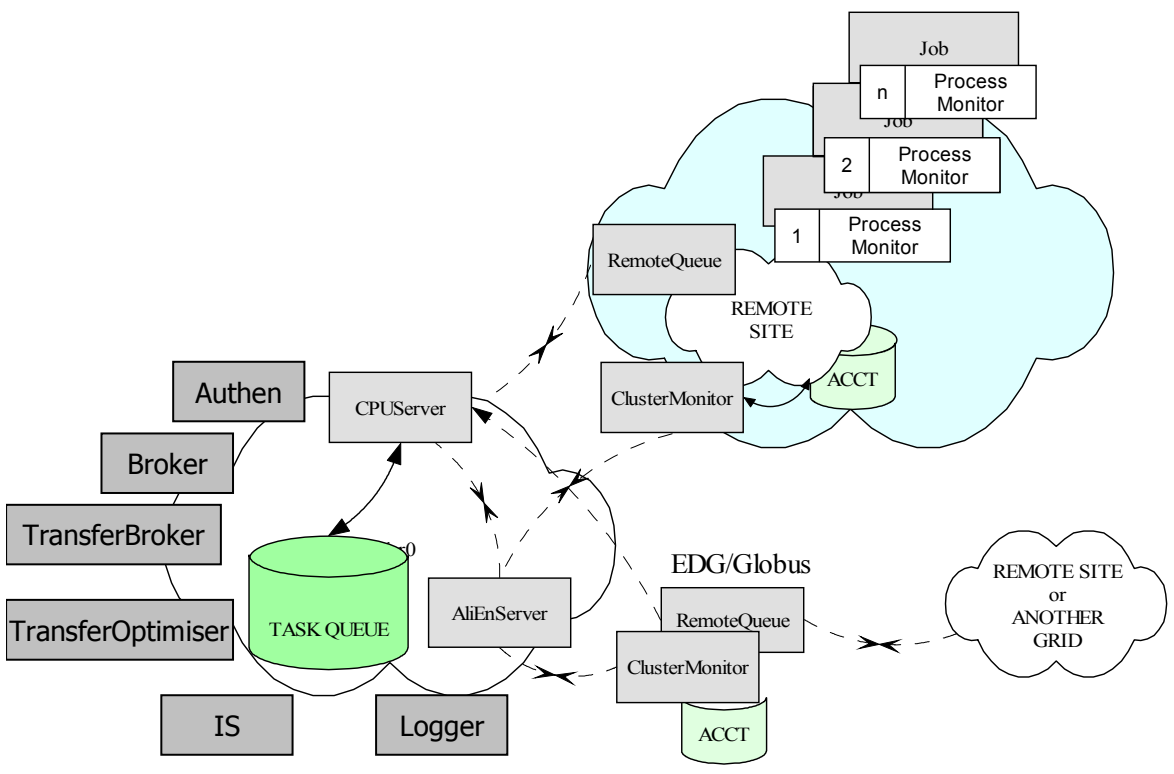

Figure 5: The job execution model

As opposed to the traditional push architecture, AliEn uses a pull model (Figure 5). In a push system, the broker knows the status of all the resources in the Grid, and it submits the job to the optimal Computing Element (CE) at a given point in time. In a pull implementation, the Broker does not need to maintain the status of the system. Instead, the remote Computing Elements monitor local resources and, once they have some shares available, they advertise themselves to the central service (CPU Server) by presenting their own ClassAds. The CPU Server will carry out the ClassAd matching against the descriptions of all tasks in the queue taking into account overall priorities and policies and, if any matching task is found, it will be given to the CE for execution. This results in a more robust and fault tolerant system, as resources can come and go. However, in order to avoid being blocked for an extended period of time, it is important that at least some resources (Tier 1 Centres) maintain reasonable quality of service that will guarantee the access to locally stored datasets.

Given the weak coupling between resources and Resource Brokers, it is possible to wrap up an entire foreign Grid as an AliEn Computing and Storage Element [3].

### 2.4.4. Computing Element

The Computing Element (CE) is an interface to the local batch system. At present, AliEn provides interfaces to LSF, PBS, BQS, DQS, Globus and Condor. The task of a CE is to get hold of jobs JDLs from the CPU Server, translate them to the syntax appropriate for the local batch system syntax and execute them. Each job is wrapped up in another Web Service (Process Monitor) allowing users to interact with the running job (send a signal or inspect the output). Prior to job execution, the CE can automatically install the software packages required by the job using the Package Manager functionality (see 2.3.5).

### 2.4.5. Storage Element

The Storage Element (SE) is responsible for saving and retrieving the files to and from the local storage. It manages disk space for files and maintains the cache for temporary files.

### 2.4.6. File Transfer

This service typically runs on the same host as the Storage Element and provides the scheduled file transfer functionality. The File Transfer Daemons (FTD) are mutually authenticated using certificates and will perform file transfer on user's behalf using the bbftp [14] protocol. File transfers are requested and scheduled in exactly the same way as jobs, this time under the control of the File Transfer Broker.

### 2.4.7. Optimizers

While the jobs or file transfer requests are waiting in the task queue, the Job and Transfer Optimizers will inspect JDLs and try to fulfil requests and resolve conflicts. This can result in the system triggering file replication in order to make job eligible to run on some sites to balance the overall load on the system. Along the same lines, one can also implement policy monitors to enforce VO policies by altering job priorities.

### 2.4.8. Logger

A Logger Service provides the mechanism for all services to report their status and error conditions. This allows Grid manager to monitor all exceptions in the system and to take corrective action.

### 2.4.9. Monitoring

The Resource Brokers do not require overall monitoring information to carry out job or file transfer scheduling. However, if available, this information would be useful to various Optimizer services. We are currently planning to deploy the MonaLisa [15] monitoring framework as a part of the AliEn Monitor Module. It will collect the monitoring information and publish it via Web Service for use by AliEn Optimizers or for visualization purposes. On the longer term, the intention is to re-use the network simulation code originally developed for MONARC [16] and now part of MonaLisa and extend it to cover the behaviour of the distributed Web of services that constitute the AliEn Grid. With this, we should be able to optimize and understand the performance of the system and verify its scalability.

## 2.5. Interfaces

### 2.5.1. Command Line Interface and GUI

The File Catalogue provides a command line interface similar to a UNIX file system with the most common commands implemented. Similar functionality is provided by the graphical user interface. Using these interfaces, it is possible to access the catalogue, submit jobs and retrieve the output.

### 2.5.2. Web Portal

AliEn provides a web portal as an alternative user interface where one can check the status of the current and past jobs, submit new jobs and interact with them. The web portal offers additional functionality to 'power' users – Grid administrators can check the status of all services, monitor, start and stop them while VO administrators (production user) can submit and manipulate bulk jobs.

### 2.5.3. C/C++ API

In order to gain access to AliEn at the application level, users need an API. Besides the perl module that allows user to use, modify or extend the interface, AliEn provides C and C++ API. In particular, the C++ API is thread safe and we used it to implement a fully featured file system on top of AliEn File Catalogue as well as an implementation of the TGrid class in ROOT to address analysis use case (see Section 3).

### 2.5.4. AliEn File System

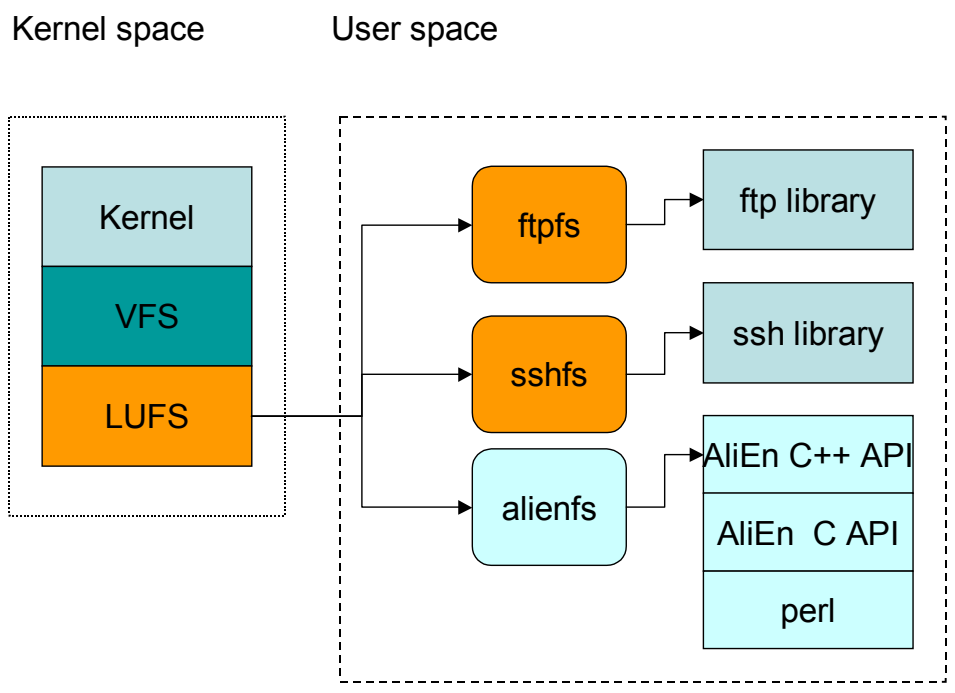

Figure 6: LUFS [17] ('Linux Userland File System')

The open source project LUFS [17] ('Linux Userland File System') offers a modular file system plug-in architecture with a generic LUFS Virtual File System (VFS) kernel module, which communicates with various user space file system modules. An implementation of LUFS based on AliEn File Catalogue (*alienfs*) has been built using the thread safe AliEn C++ API [18]. To use such file system, each user has to execute a *mount* command and authenticate to the Grid. The authentication is done only once for the user space thread and, in case the connection is broken, it is automatically restored when required. Therefore it is recommended to use a SSH key or proxy-certificate based authentication method.

To provide even more flexible framework, a general *gridfs* module has been developed. This module allows dynamic loading of Grid API libraries and enables to use the same LUFS module for various Grid platforms [18].

While scheduled transfer and file replication as described in Section 2.4.6. are sufficient to fulfil the requirements of distributed batch style processing, interactive analysis and applications like the file system require instant and efficient response in addition to Grid security.

Since analysis datasets are shared between potentially large communities of users, it is reasonable to move away from a point-to-point connection scheme between Storage Elements and user application towards connections of a cache-gateway type. In this model, application file access is routed through distributed cache servers to allow efficient usage of network resources. A more detailed description of the AliEn File System and the CrossLink Cache I/O architecture can be found in [18].

## 3. ANALYSIS WITH ALIEN & ROOT

As the next generation of physics experiments expects to produce data sets of the order of PBytes per year, conventional data processing will not be anymore sufficient to extract the relevant physics data out of such enormous datasets. The expected volume of data which needs to be handled as well as the nature of worldwide collaboration necessitates a distributed approach where data and computing will be geographically distributed. While the existing processing schemes using batch queues and mass storage systems have proven in the past that large datasets can be handled very well inside individual computer centres, scaling this model to the global scale presents new and formidable computing challenges.

AliEn can be used today to solve typical HEP use cases like running simulation, reconstruction and subsequent centrally coordinated processing steps as it was outlined above. It is likely that a large fraction of the CPU time use in HEP will actually be spent in executing these steps. However, end users will be waiting for output of these organized production steps to start a largely chaotic process of analysis where many of them (hundreds) will be trying to run their (possibly private) algorithms over a large subset of distributed datasets with the aim to extract the physics observables specific to their analysis.

In the following section we will describe how AliEn can be coupled with ROOT to provide an analysis platform suitable for typical HEP experiments.

## 3.1. AliEn as Analysis Platform

AliEn provides the two key elements important for large-scale distributed data processing – a global file system for data storage and the possibility to execute jobs in a distributed environment, harvesting available CPU resources in an optimal way. In addition, while solving the HEP analysis use case we can take advantage of trivial parallelization that can be achieved at the 'physics event' level. In the remaining part of this section an outline of possible solutions of the analysis use case using the combination of AliEn and ROOT will be given.

### 3.1.1. An Outline Analysis

As the first step, the analysis framework has to extract a subset of the datasets from the virtual file catalogue using metadata conditions provided by the user. The next and the most difficult part is the splitting of the tasks according to the location of data sets. A trade-off has to be found between best use of available resources and minimal data movements. Ideally jobs should be executed where the data are stored. Since one cannot expect a uniform storage location distribution for every subset of data, the analysis framework has to negotiate with dedicated Grid services the balancing between local data access and data replication.

Once the distribution is decided, the analysis framework spawns sub-jobs. These are submitted to the Resource Broker with precise job descriptions. The user can control the results while and after data are processed. The framework collects and merges available results from all terminated sub-jobs on request.

It is desirable that once an analysis task is defined, the corresponding analysis objects become persistent in the Grid environment so the user can go offline and reload an analysis task at a later date, check the status, merge current results or resubmit the same task with modified analysis code.

### 3.1.2. Adding Grid functionality to ROOT

The analysis framework used by ALICE and by most of current HEP experiments is ROOT. To enable analysis on the Grid using AliEn, several new classes were required.

The *TAlien* class, based on the abstract *TGrid* class, implements the basic methods to connect and disconnect from the Grid environment and to browse the virtual file catalogue. *TAlien* uses the AliEn API for accessing and browsing the file catalogue.

The files are handled in ROOT by the *TFile* class, which provides a plug-in mechanism supporting various file access protocols. The *TAlienFile* class inherits from the *TFile* class and provides additional file access protocols using the generic file access interface of the AliEn C++ API.

To submit jobs to AliEn from ROOT we need a representation of a queue job object. *TAlienJob class* encapsulates all the characteristics of an AliEn queue job like:

- Job Identifier
- Job Requirements (JDL)
- Job Status

The *TAlienJobIO* class is designed to allow transparent job processing of several files. The class opens sequentially files given in an I/O configuration file that specifies locations and access methods.

Finally, the *TAlienAnalysis* class provides the overall steering as described in Section 3.1.1.

The user has to provide the following information for an analysis task:

- Tag name for an analysis object.
- Name of the macro to execute.
- Name of the interpreter to use for macro execution (in general this is ROOT, for the ALICE experiment it is AliROOT).
- Necessary input files for the macro (like calibration tables, cut configuration files etc.).
- Data set selection criteria
    - Top level directory in the virtual file catalogue to start metadata matching (virtual directories can be considered already as a pre-selection).
    - Meta data selection criteria.
- Splitting hints
    - Number of sub-jobs to spawn (considered only as a guideline).
    - Job distribution level (keeps jobs possibly together in one site or disperses to as many as possible sites).
- List of output files produced by each sub-job.

In the next step the analysis class selects a data set from the virtual file catalogue using the input parameters and generates JDL and JIO (Job Input Objects) files for each sub-job. The JDL file contains the sub-job requirements like input and output sandbox, resource needs such as storage and/or computing elements. The JIO file contains a list of input objects describing the data files to process and the mechanisms to access these files.

Once instantiated, the *TAlienAnalysis* object creates a subdirectory in the user home directory in the AliEn File Catalogue with the given tag name and registers all JDL and JIO files in the same directory. The analysis object itself is written into a root file that is again registered inside the same directory.

Once the analysis is running, the user can query the status of the spawned sub-jobs and start to retrieve results from the sub-jobs. The ROOT output of each sub-job can be found in the standard AliEn job output directory and is brought automatically onto the user's machine. The analysis class merges all sub-job ROOT files. Histograms are summed and ROOT trees are merged into chain objects. Once all sub-jobs are executed, the final result is registered in the file catalogue.

The analysis concept as presented can be considered more as a batch analysis with interactive capability. Since sub-jobs are completely independent from each other and run asynchronously, an analysis object can be created interactively at the ROOT prompt or by submitting a job to the Grid.

To boost the interactive aspect of the model, the user should spawn many small jobs in order to start receiving back results very fast: while a small fraction of the jobs will be executed immediately, the remaining sub-jobs will wait in the queue to be executed sequentially when the previous bunch of jobs has finished and results are made available.

### 3.1.3. Interactive Analysis Facility with PROOF and AliEn

PROOF [19] - the parallel ROOT facility - covers the needs of interactive analysis in local cluster environment. To enable the use of PROOF in a Grid environment, two changes are necessary:

- Create a hierarchy of PROOF slave/master servers.
- Use Grid (AliEn) services to facilitate job splitting.

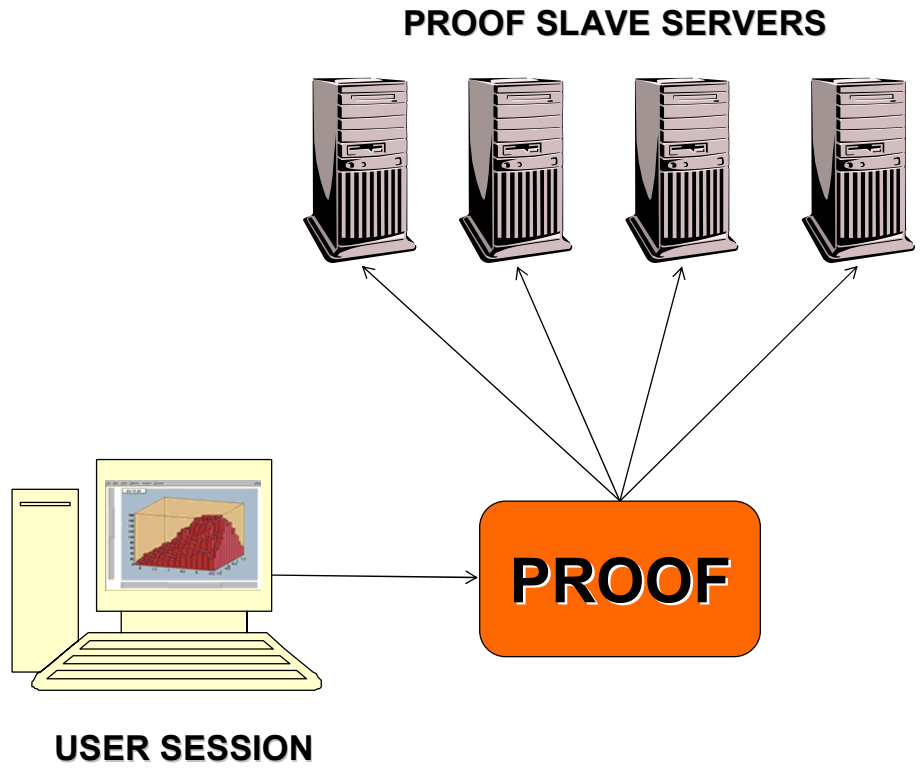

Figure 7: The conventional setup of a PROOF farm.

In the conventional setup (Figure 7), PROOF slave servers are managed by a PROOF master server, which distributes tasks and collects results. In a multi-site setup each site running a PROOF environment will be seen as a SuperPROOF slave server for a SuperPROOF master server running on the user machine. The PROOF master server has therefore to implement the functionality of a PROOF master server and a SuperPROOF worker server at the same time. AliEn classes used for asynchronous analysis as described in Section 3.1.2 can be used for task splitting in order to provide the input data sets for each site that runs PROOF locally.

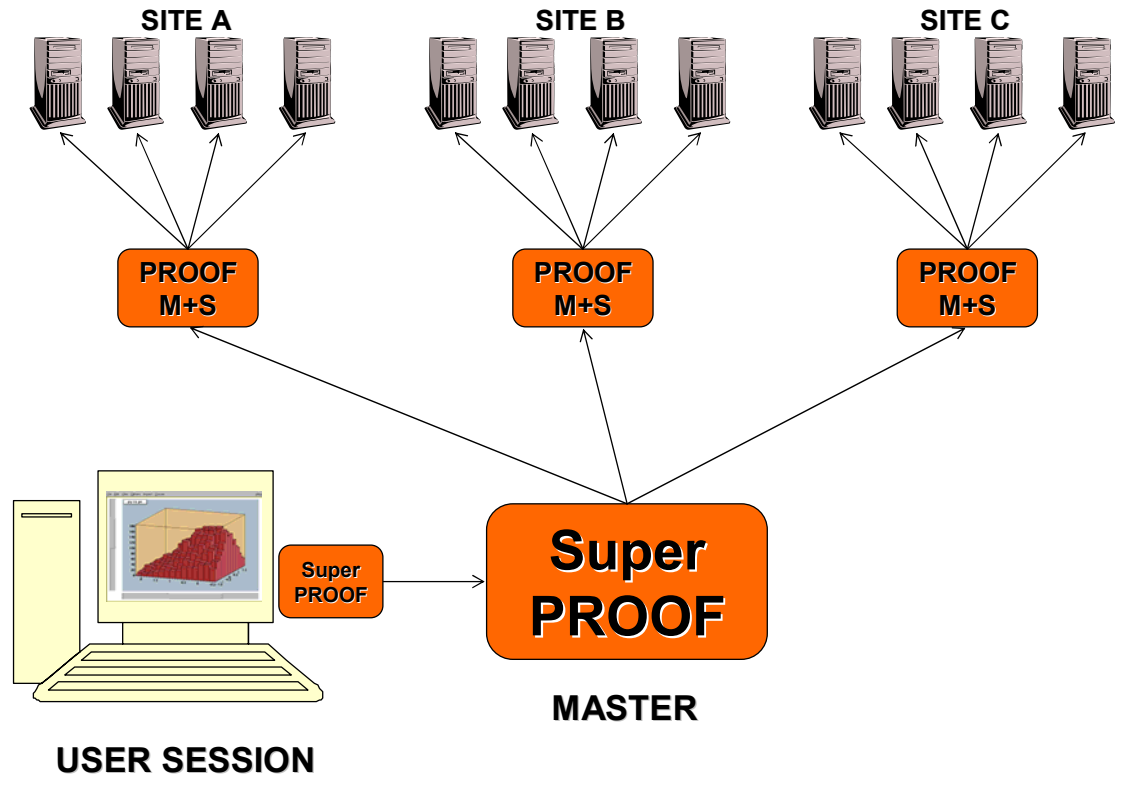

Figure 8: SuperPROOF – multi-site PROOF setup.

The SuperPROOF master server assigns these data sets to the PROOF master server on individual sites. Since these data sets are locally readable by all PROOF servers, the PROOF master on each site can distribute the data sets to the PROOF slaves in the same way like for a conventional setup (Figure 8).

### 3.1.4. Enabling SuperPROOF using AliEn

In a static scenario, each site maintains a static PROOF environment. To become a PROOF master or slave server a farm has to run a *proofd* process. These daemons are always running on dedicated machines on each site. To start a SuperPROOF session the SuperPROOF master contacts each PROOF master on individual sites, which start PROOF workers on the configured nodes. The control of the *proofd* processes can be responsibility of the local site or it can be done by a dedicated AliEn service.

In a dynamic environment, once a SuperPROOF session is started, an AliEn Grid service starts *proofd* processes dynamically using dedicated queues in the site batch queue system. This is assuring that a minimum set of *proofd* processes is always running. The system can react to an increasing number of requests for SuperPROOF sessions with starting a higher number of *proofd* processes.

The static environment makes sense for sites with large computing capabilities, where farm nodes can be dedicated exclusively to PROOF. If a site requires an efficient resource sharing, a dynamic environment appears to be the best choice, since it can use locally configured job queues to run *proofd* processes. These processes can be killed, if the requests decrease or after a period of inactivity.

The SuperPROOF concept can have an important impact during the development of the analysis algorithms. The interactive response will simplify the work of the physicist when processing large data sets and it will result in a higher efficiency while prototyping the analysis code. On the other hand the stateless layout of parallel batch analysis with ROOT will be the method of choice to process complete data sets with final algorithms. Scheduled jobs allow for a more efficient use of resources since the system load is predictable from the JDLs of queued jobs. In general, the fault tolerance of a highly synchronized system like SuperPROOF with unpredictable user behaviour is expected to be lower than the one of an asynchronous system. While the prototype of analysis system based on AliEn already exists, the first prototype of SuperPROOF is foreseen for the end of year 2003.

## 4. OUTLOOK

### 4.1. AliEn and OGSA

Open Grid Services Architecture [20] (OGSA) has been proposed as a common foundation for future Grids. AliEn was conceived and developed from the beginning using the same basic technology as proposed by OGSA. This makes it possible to implement smoothly new standards as soon as they become commonly accepted. As an interim solution, the MammoGrid [21] project will explore a possibility to create the OGSA compatible front-end Grid Services as a proxy to AliEn Web Services.

### 4.2. Virtual server

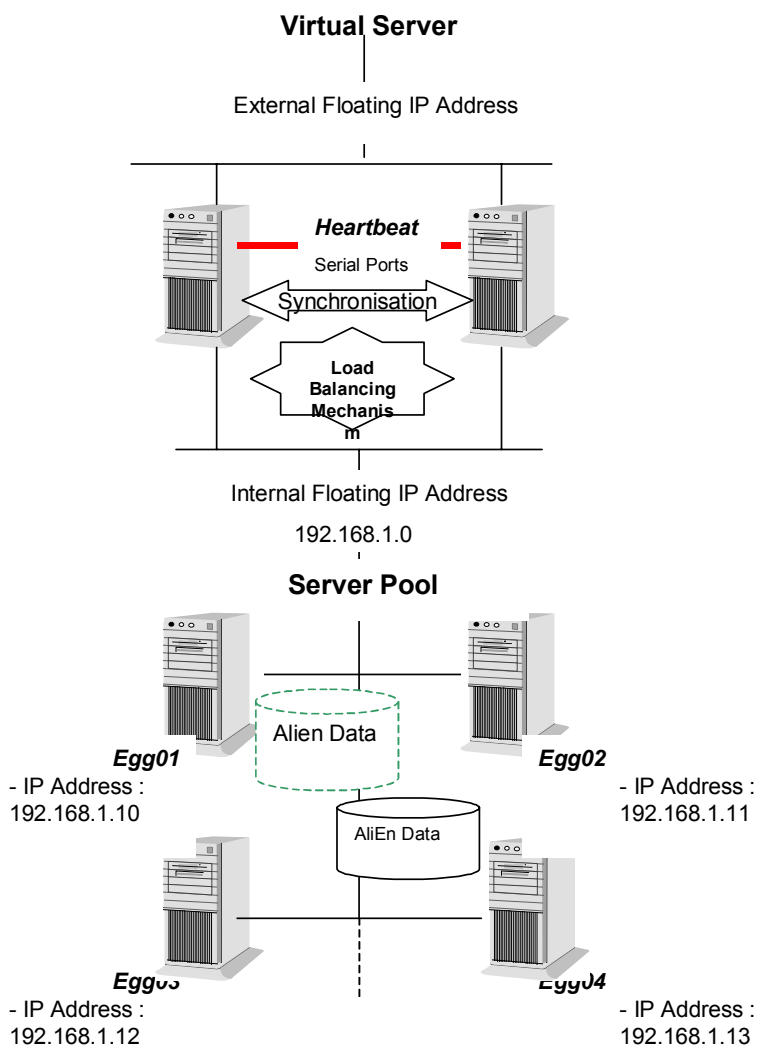

Figure 9: AliEn Virtual Server

One of the practical problems we have encountered while setting up the AliEn Grid service is reliability and scalability of a central service that supports the entire Virtual Organization (authentication service, database). The problem is further complicated by the need to maintain the configuration and run the services for many Virtual Organizations while still debugging and developing the system. The solution we found was to setup a high availability cluster with redundant servers running the services for different Virtual Organizations and use generic front end services that can do dynamic and configurable routing of SOAP/XML messages (Figure 9).

### 4.3. The Federation of Grids

Due to weak coupling between the resources and the Resource Brokers in the AliEn Grid model it is possible to imagine a hierarchical Grid structure that spans multiple AliEn and "foreign" Grids but also includes all resources under the direct control of top level Virtual Organization. The connectivity lines in Figure 10 represent the collaboration and trust relationships. In this picture the entire foreign Grid can be represented as a single Computing and Storage Element (albeit a potentially powerful one). In this sense, we have constructed the AliEn-EDG interface and tested the interoperability [3]. Along the same lines, AliEn-AliEn interface allows creation of federation of collaborating Grids. The resources in this picture can be still shared between various top level Virtual Organizations according to the local site policy so that the Grid federations can overlap at resource level.

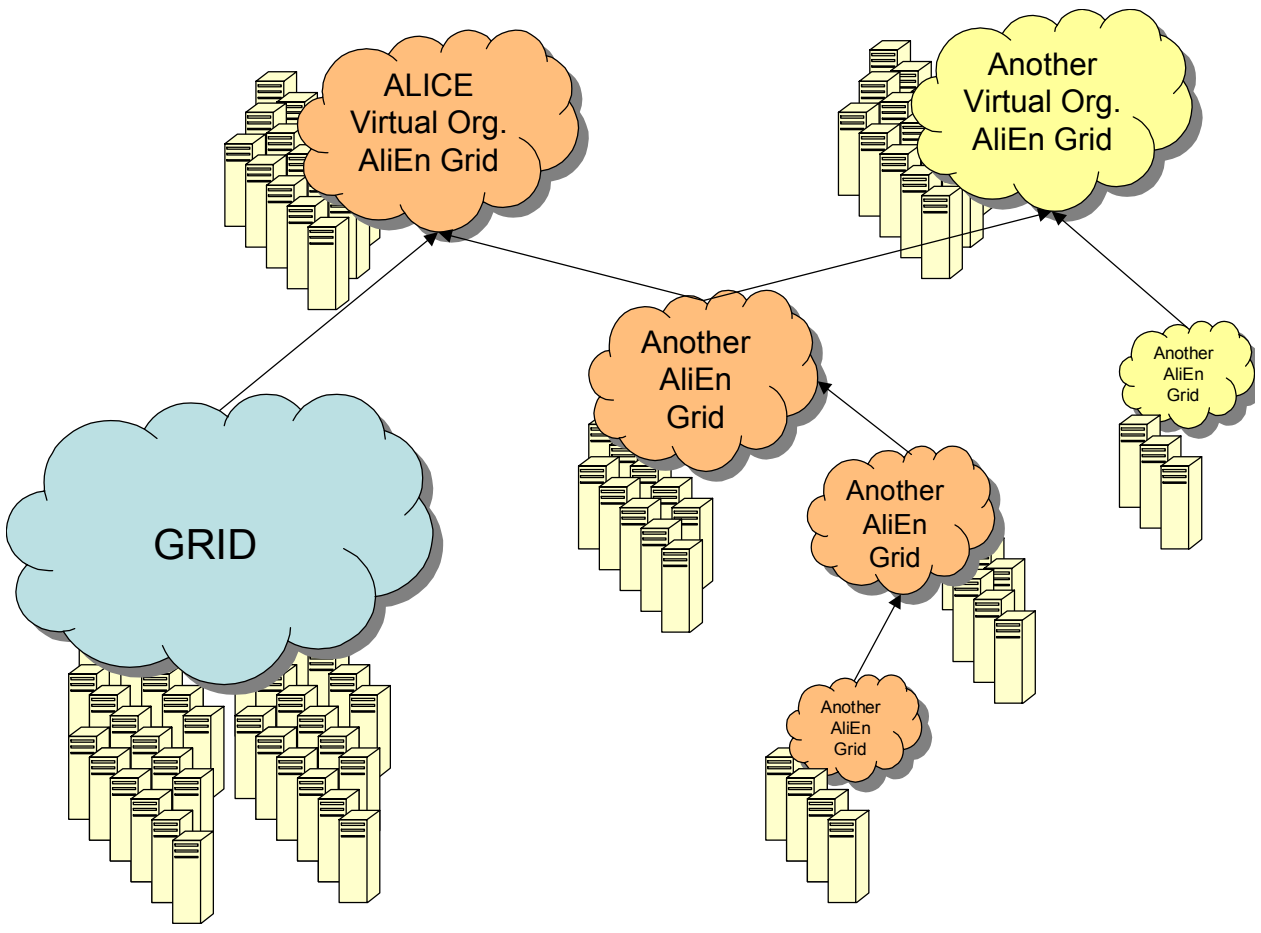

Figure 10: The federation of collaborating Grids

However, when it comes to implementing data Grids suitable for HEP, this kind of flexibility does not come without price – each Virtual Organization must maintain its own File and Metadata Catalogue.

### 4.4. Other projects using AliEn

Besides the ALICE experiment, several HEP experiments and medical projects have expressed their interest in AliEn or some components of it. In particular, AliEn is used to provide the Grid component for MammoGRID [21] and GPCALMA [22].

## 5. CONCLUSIONS

AliEn has been tested in several large-scale productions. In the first one, during November 2001, almost six thousand Pb+Pb events were generated. To date, many production rounds have been completed and the number of collaborating sites has increased to more than 30. There were up to 450 concurrently running jobs in recent productions and overall more than 25TB of data has been generated.

In its current state, AliEn provides sufficient functionality to carry out typical simulation and reconstruction tasks. The distributed analysis starting from the ROOT prompt can be handled by transparently decomposing a complex request into several sub-jobs, executing them and presenting the output back to the user. A more advanced and elaborate environment that will include joining several PROOF clusters into a SuperPROOF cluster using the AliEn infrastructure is currently under development.

## Acknowledgments

The authors wish to thank Hewlett-Packard for supplying the high availability servers needed to run the central AliEn services.